\documentclass[]{article}

\usepackage[top = 0.75in, right = 0.75in, left = 0.75in, bottom = 0.75in]{geometry}
\usepackage{authblk}
\usepackage{siunitx}
\usepackage{amsmath}
\usepackage{mathtools}
\usepackage{mathrsfs}
\usepackage{amssymb}
\usepackage{graphicx}
\usepackage{textcomp}
\usepackage{epstopdf}
\usepackage{afterpage}
\usepackage{tabulary}
\usepackage{rotating}
\usepackage{multirow}
\usepackage{longtable}
\usepackage{amsthm}
\usepackage{stmaryrd}
\usepackage{url}
\usepackage{float}
\usepackage{tablefootnote}

\title{A Dynamic Parametric Wind Farm Model for Simulating Time-varying Wind Conditions and Floating Platform Motion}
\author[1]{Ali C. Kheirabadi}
\author[1]{Ryozo Nagamune}
\affil{The University of British Columbia, Vancouver Campus, 2054-6250 Applied Science Lane, Vancouver, BC Canada V6T 1Z4}

\begin{document}

\maketitle

\begin{abstract}
	
This paper introduces a dynamic parametric wind farm model that is capable of simulating floating wind turbine platform motion coupled with wake transport under time-varying wind conditions. The simulator is named FOWFSim-Dyn as it is a dynamic extension of the previously developed steady-state \textit{Floating Offshore Wind Farm Simulator} (FOWFSim). One-dimensional momentum conservation is used to model dynamic propagation of wake centerline locations and average velocities, while momentum recovery is approximated with the assumption of a constant temporal wake expansion rate. Platform dynamics are captured by treating a floating offshore wind farm as a distribution of particles that are subject to aerodynamic, hydrodynamic, and mooring line forces. The finite difference method is used to discretize the momentum conservation equations to yield a nonlinear state-space model. Simulated data are validated against steady-state experimental wind tunnel results obtained from the literature. Predictions of wake centerlines differed from experimental results by at most $8.19\,\si{\percent}$ of the rotor diameter. Simulated wake velocity profiles in the far-wake region differed from experimental measurements by less than $3.87\,\si{\percent}$ of the free stream wind speed. FOWFSim-Dyn thus possesses a satisfactory level of fidelity for engineering applications. Finally, dynamic simulations are conducted to ensure that time-varying predictions match physical expectations and intuition.

\end{abstract}

\section{Introduction}

Since the introduction of parametric wake models by Jensen~\cite{Jensen1983} and Kati\'c \textit{et al.}~\cite{Katic1986}, such wind farm simulators have served as essential tools for enhancing wind farm performance. This enhancement has been achieved via two distinct fields of study. The older of the two is \textit{layout optimization}, wherein the optimal installation locations of wind turbines are computed with the objective of maximizing annual revenue~\cite{Shakoor2016}. Since such optimization problems are solved offline prior to wind farm construction, steady-state wake models have sufficed for estimating annual energy production. The field of study that has more recently experienced a surge in interest is \textit{wind farm control}, which involves real-time wind turbine actuation for the purpose of manipulating the wind field to achieve some wind farm-level objective~\cite{Kheirabadi2019a}. This ultimate goal may be efficiency maximization, or power output tracking with turbine load alleviation~\cite{Knudsen2015}. In either case, since actuators are adjusted in real-time, dynamic wake phenomena such as turbulence, transport delay, time-varying mean wind speed and direction, and floating platform motion (for deep-water offshore wind farms) are pertinent when evaluating controller performance.

Steady parametric wake models have been used successfully to raise wind farm efficiency in large eddy simulations (LES)~\cite{Gebraad2016} and field tests~\cite{Fleming2017}. Further, in one instance, Gebraad \textit{et al.}~\cite{Gebraad2015} reported no significant performance gains when using a dynamic wake model for wind farm control in contrast to using a steady wake model. Nonetheless, there are benefits associated with using dynamic wake models. First, traditional state and parameter estimation techniques may be used to adapt such models to time-varying wind conditions~\cite{Gebraad2015}. Second, low-fidelity dynamic wake models may be used to test controller robustness against time-varying wind conditions prior to dedicating time and resources to conducting high-fidelity simulations and field tests (as performed by Johnson and Fritsch~\cite{Johnson2012} and Gebraad and van Wingerden~\cite{Gebraad2015b}). Further research comparing wind farm control based on dynamic versus steady parametric models may reveal additional benefits. Such progress will only be possible, however, provided the availability of various dynamic wake models.

Reviews of wake modeling may be found in the works of Boersma \textit{et al.}~\cite{Boersma2017}, G\"o\c{c}men \textit{et al.}~\cite{Gocmen2016}, Vermeer \textit{et al.}~\cite{Vermeer2003}, and in our previous review article~\cite{Kheirabadi2019a}. We will focus our current discussion on parametric dynamic wake models. The earliest of such models found application in power de-rating wind farm control research conducted by Gebraad and van Windgerden~\cite{Gebraad2015b}, Johnson and Fritsch~\cite{Johnson2012}, and Ahmad \textit{et al.}~\cite{Ahmad2014}. Power de-rating involves reducing the thrust force exerted onto the wind by upstream wind turbines as a means of increasing the fluid momentum available to downstream machines~\cite{Johnson2009}. Since this application involves neither wake deflection nor wind turbine relocation, wake dynamics in these studies were modeled using time-delays in computed steady-state incident wind speeds. These time-delays represented the duration required for changes in the wind field at some upstream turbine to propagate to downstream machines. This approach is valid as long as the wind direction remains constant, and wake centerline deflection and turbine relocation are not pertinent.

In order to account for transport delay of steered wakes, Gebraad and van Wingerden~\cite{Gebraad2014b} developed the Flow Redirection and Induction Dynamics (FLORIDyn) model. Their approach involved representing flow within a wake using translating points that were initialized at each turbine and then transported downstream. Each point contained information regarding its corresponding turbine's operating parameters at the instant in time at which the point was initialized. Using this information, wake properties at the downstream location of the translating point were obtained using the Flow Redirection and Induction in Steady-state (FLORIS) wake model~\cite{Gebraad2016}, which utilizes integral forms of mass and momentum conservation to compute downstream wake properties. More simply put, FLORIDyn transports steady-state wake characteristics computed with FLORIS in the free stream wind direction. Time varying wind direction and floating platform motion are not considered in FLORIDyn however.

In an alternative approach, Shapiro \textit{et al.}~\cite{Shapiro2017} used the differential forms of mass and momentum conservation to simulate dynamic wake behavior. Local and convective wake accelerations in the free stream wind direction were described by material derivatives, and these accelerations were equated to force terms representing turbulent mixing and rotor thrust. The advantage of this modeling approach was that wake transport would be inherently captured by convective acceleration terms, thus eliminating the need for the translating points employed by the FLORIDyn model. Instead, all wake characteristics were functions of a fixed grid in the downstream direction. Shapiro \textit{et al.}~\cite{Shapiro2018a} later extended their model to capture wake redirection resulting from rotor yaw misalignment. Prandtl's lifting line theory was used to compute transverse wake velocities, shed circulation, and vortex properties immediately downstream of yawed rotors. Wake centerline deflection in the free stream wind direction was then computed by equating the material derivative of the centerline position to the transverse component of the wake velocity. These works do not capture floating platform motion or time-varying wind speed and direction.

Finally, Boersma \textit{et al.}~\cite{Boersma2018} developed the \textit{Wind Farm Simulator} (WFSim), which is a control-oriented dynamic wake model based on the two-dimensional form of the unsteady turbulent Navier-Stokes equations. The major benefit of WFSim is that individual wake expansion and the interaction of multiple wakes are inherently captured by the mixing length turbulence model employed. In the previously discussed models, the rate of linear wake expansion was either estimated or assumed. Further, the previous models simulated flow behavior in regions with overlapping wakes by assuming that the effective kinetic energy deficit in the wind field is equal to the sum of deficits corresponding to all pertinent wakes. Despite its higher-fidelity, WFSim requires approximately $1000\,\si{\sec}$ of computation for $1000\,\si{\sec}$ of simulation in comparison to previously discussed models (several seconds of computation for $1000\,\si{\sec}$ of simulation). Moreover, WFSim does not model floating platform motion or time-varying wind speed and direction.

In the current paper, we loosely follow the approach of Shapiro \textit{et al.}~\cite{Shapiro2017}, whereby partial differential equations are used to capture wake transport, and we develop a dynamic parametric wake model capable of simulating time-varying wind speed and direction, along with platform motion for floating offshore wind farms. The novelty of this paper therefore includes the following: (i) additional terms in the wake momentum conservation equations to capture time-varying free stream wind velocity effects; and (ii) a coupled dynamic model that captures planar floating wind turbine motion in the presence of aerodynamic interaction. Our approach is physics-based with the rate of wake expansion as the only parametric assumption. This model serves as a dynamic extension of our previously developed steady-state tool~\cite{Kheirabadi2020}, which was named the \textit{Floating Offshore Wind Farm Simulator} (FOWFSim), and will henceforth be referred to as FOWFSim-Dyn. Fixed-foundation wind farms may also be modeled by simply deactivating turbine platform motion.

The remainder of this paper is organized as follows: Section~\ref{Section - Mathematical model} provides a detailed mathematical description of FOWFSim-Dyn along with a discussion of its limitations. In Section~\ref{Section - Model tuning and validation}, we perform a mesh convergence study and validate FOWFSim-Dyn using steady-state experimental results reported by Bastankhah and Port\'e Agel~\cite{Bastankhah2016}. We also present dynamic simulation results that demonstrate the various capabilities of FOWFSim-Dyn. We finally conclude the paper in Section~\ref{Section - Conclusions and recommendations} by listing potential research directions for enhancing FOWFSim-Dyn.

\section{Mathematical model - FOWFSim-Dyn} \label{Section - Mathematical model}

This section details the mathematical formulation behind FOWFSim-Dyn. First, the problem setup, solver block diagram, and resulting equations of motion are presented in Sections~\ref{Subsection - Wind farm description}--\ref{Subsection - Wind interaction model}. Finally, important assumptions and limitations pertaining to FOWFSim-Dyn are discussed in Section~\ref{Subsection - Assumptions and limitations}.

\subsection{Wind farm description} \label{Subsection - Wind farm description}

\begin{figure}
	\centering
	\includegraphics[width=4in]{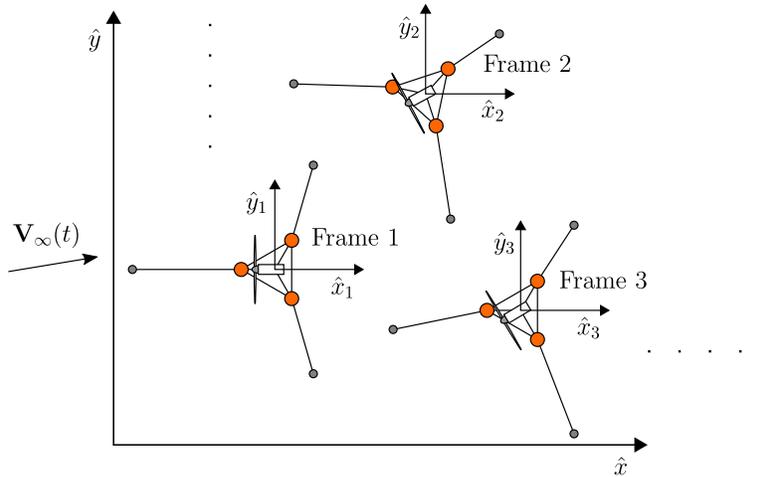}
	\caption{Schematic of a general floating offshore wind farm with semi-submersible platforms used as a basis for FOWFSim-Dyn's mathematical model.} \label{Figure - Dyn model - Wind farm description}
\end{figure}

Figure~\ref{Figure - Dyn model - Wind farm description} shows a top view schematic of the general floating offshore wind farm that we model in the current work. Floating wind turbines are treated as a system of particles that are distributed along the two-dimensional ocean surface. Throughout this paper, we consider only three-cylinder semi-submersible floating platforms as per the baseline design presented by Robertson \textit{et al.}~\cite{Robertson2014}. Each floating structure is therefore connected to three anchors via mooring lines for the purpose of station-keeping. % Further, the turbines are based on the baseline $5\,\si{MW}$ design presented by Jonkman \textit{et al.}~\cite{Jonkman2009}. Details of the platform and turbine designs are listed in Appendix~XXX.

We define the set $\mathscr{F} = \left\{ 1, 2, \cdots, N \right\}$ to denote the indices of the $N$ floating wind turbines within the wind farm, and we refer to each individual turbine using the identifier $i$. We then number the wind turbines in ascending order based on their downstream location. That is to say, the most upstream turbine is numbered by $i = 1$, while the most downstream machine is identified by $i = N$.

The fixed global frame of reference is identified by the $\hat{x}$ and $\hat{y}$ axes. Each wind turbine also possesses a local non-inertial translating (though not rotating) reference frame that is attached to its center of gravity. We identify the reference frame that is fixed to turbine~$i$ as frame~$i$. Further, the axes of frame~$i$ are referred to as $\hat{x}_i$ and $\hat{y}_i$.

We assume that a predominant wind direction exists, and that it is aligned with the positive $\hat{x}$ axis. The free stream wind velocity is then denoted by the vector $\mathbf{V}_\infty(t)$, which contains $\hat{x}$ and $\hat{y}$ components $U_\infty(t)$ and $V_\infty(t)$ as follows:
\begin{equation}
	\mathbf{V}_\infty(t) \coloneqq
	\begin{bmatrix}
		U_\infty(t) & V_\infty(t)
	\end{bmatrix}^\mathrm{T}.
\end{equation}
$U_\infty(t)$ therefore represents the free stream wind speed in the predominant wind direction, while $V_\infty(t)$ accounts for fluctuations in the transverse free stream wind speed.

\subsection{States and inputs}

Ultimately, FOWFSim-Dyn takes the following nonlinear state-space form:
\begin{equation}
	\dot{\mathbf{x}}_\mathrm{farm}(t) \coloneqq f(\mathbf{x}_\mathrm{farm}(t), \mathbf{u}_\mathrm{farm}(t), \mathbf{V}_\infty(t)),
\end{equation}
where the wind farm state vector $\mathbf{x}_\mathrm{farm}(t)$ combines the floating wind turbine state vector $\mathbf{x}(t)$ with the wake state vector $\mathbf{x}_\mathrm{w}(t)$ as follows:
\begin{equation}
	\mathbf{x}_\mathrm{farm}(t) \coloneqq
	\begin{bmatrix}
		\mathbf{x}^\mathrm{T}(t) & \mathbf{x}_\mathrm{w}^\mathrm{T}(t)
	\end{bmatrix}^\mathrm{T}.
\end{equation}
The wind turbine state vector $\mathbf{x}(t)$ comprises the position and velocity vectors of all floating wind turbines within the wind farm as follows:
\begin{equation}
	\mathbf{x}(t) \coloneqq
	\begin{bmatrix}
		\mathbf{r}_1^\mathrm{T}(t) & \mathbf{r}_2^\mathrm{T}(t) & \cdots & \mathbf{r}_N^\mathrm{T}(t) & \mathbf{v}_1^\mathrm{T}(t) & \mathbf{v}_2^\mathrm{T}(t) & \cdots & \mathbf{v}_N^\mathrm{T}(t)
	\end{bmatrix}^\mathrm{T},
\end{equation}
where $\mathbf{r}_i(t)$ and $\mathbf{v}_i(t)$ are vectors containing $\hat{x}$ and $\hat{y}$ components of the position and velocity of turbine~$i$ as follows:
\begin{eqnarray}
	\mathbf{r}_i(t) & \coloneqq &
	\begin{bmatrix}
		x_i(t) & y_i(t)
	\end{bmatrix}^\mathrm{T}, \\
	\mathbf{v}_i(t) & \coloneqq &
	\begin{bmatrix}
		v_{x,i}(t) & v_{y,i}(t)
	\end{bmatrix}^\mathrm{T}.
\end{eqnarray}

The wake state vector $\mathbf{x}_\mathrm{w}(t)$ contains the states of the wakes generated by the $N$ floating wind turbines as follows:
\begin{equation} \label{Equation - Wake state vector}
	\mathbf{x}_\mathrm{w}(t) \coloneqq
	\begin{bmatrix}
		\mathbf{x}_{\mathrm{w},1}^\mathrm{T}(t) & \mathbf{x}_{\mathrm{w},2}^\mathrm{T}(t) & \cdots & \mathbf{x}_{\mathrm{w},N}^\mathrm{T}(t)
	\end{bmatrix}^\mathrm{T}.
\end{equation}
Assuming that the states of wake~$i$ are defined at $N_{\mathrm{p},i}$ discrete points along the downstream direction, $\mathbf{x}_{\mathrm{w},i}(t)$ comprises the states of wake~$i$ at each of these discrete points as follows:
\begin{equation} \label{Equation - Single wake state vector}
	\mathbf{x}_{\mathrm{w},i}(t) \coloneqq
	\begin{bmatrix}
		\mathbf{x}_{\mathrm{w},i,1}^\mathrm{T}(t) & \mathbf{x}_{\mathrm{w},i,2}^\mathrm{T}(t) & \cdots & \mathbf{x}_{\mathrm{w},i,N_{\mathrm{p},i}}^\mathrm{T}(t)
	\end{bmatrix}^\mathrm{T}.
\end{equation}
The state vector $\mathbf{x}_{\mathrm{w},i,p}(t)$ at each point $p$ along wake~$i$ then consists of the wake centerline location $y_{\mathrm{w},i,p}(t)$, wake velocity components $u_{\mathrm{w},i,p}(t)$ and $v_{\mathrm{w},i,p}(t)$, which correspond to the $\hat{x}_i$ and $\hat{y}_i$ directions, and the wake diameter $D_{\mathrm{w},i,p}(t)$ as follows:
\begin{equation}  \label{Equation - Wake point state vector}
	\mathbf{x}_{\mathrm{w},i,p}(t) \coloneqq
	\begin{bmatrix}
		y_{\mathrm{w},i,p}(t) & u_{\mathrm{w},i,p}(t) & v_{\mathrm{w},i,p}(t) & D_{\mathrm{w},i,p}(t)
	\end{bmatrix}^\mathrm{T}.
\end{equation}
These wake characteristics are portrayed in Fig.~\ref{Figure - Wake characteristics} and discussed in Section~\ref{Subsection - Single wake model}.

The wind farm input vector $\mathbf{u}_\mathrm{farm}(t)$ contains the input vectors for the $N$ floating wind turbines as follows:
\begin{equation}
	\mathbf{u}_\mathrm{farm}(t) \coloneqq
	\begin{bmatrix}
		\mathbf{u}_1^\mathrm{T}(t) & \mathbf{u}_2^\mathrm{T}(t) & \cdots & \mathbf{u}_N^\mathrm{T}(t)
	\end{bmatrix}^\mathrm{T},
\end{equation}
where $\mathbf{u}_i(t)$ consists of the axial induction factor $a_i(t)$ and yaw angle $\gamma_i(t)$ of turbine~$i$ as follows:
\begin{equation}
	\mathbf{u}_i(t) \coloneqq
	\begin{bmatrix}
		a_i(t) & \gamma_i(t)
	\end{bmatrix}^\mathrm{T},
\end{equation}
with all yaw angles defined as positive counter-clockwise from the $\hat{x}$ axis.

\subsection{Solver block diagram} \label{Subsection - FOWFSim block diagram}

\begin{figure}
	\centering
	\includegraphics[width=4in]{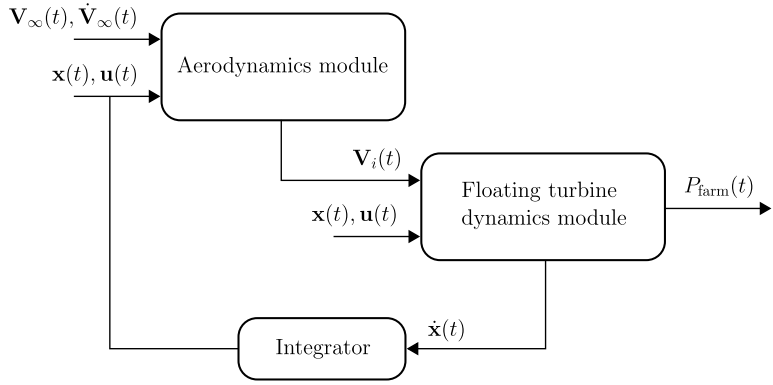}
	\caption{Block diagram showing the computation modules of FOWFSim-Dyn along with information transfer routes.} \label{Figure - Dynamic - FOWFSim block diagram}
\end{figure}

The block diagram for FOWFSim-Dyn is shown in Fig.~\ref{Figure - Dynamic - FOWFSim block diagram}. The simulator consists of two main modules. The \textit{aerodynamics module} requires the states $\mathbf{x}(t)$ and inputs $\mathbf{u}(t)$ of all turbines, along with the free stream wind velocity and acceleration vectors $\mathbf{V}_\infty(t)$ and $\dot{\mathbf{V}}_\infty(t)$ at time $t$. Its function is to compute the effective wind velocity vector $\mathbf{V}_i(t)$ that is incident on the rotor of turbine~$i$ for all $i \in \mathcal{F}$.

The \textit{Floating turbine dynamics module} uses these incident wind velocity vectors, along with turbine states and inputs, to compute the rates of change of turbine states $\dot{\mathbf{x}}_i(t)$ at time $t$. Using a standard ordinary differential equation solver, state derivatives are integrated to compute state trajectories over time. This module also computes the power outputs of individual wind turbines as well as that of the entire wind farm $P_\mathrm{farm}(t)$.

\subsection{Wind farm power output}

\begin{figure}
	\centering
	\includegraphics[width=3in]{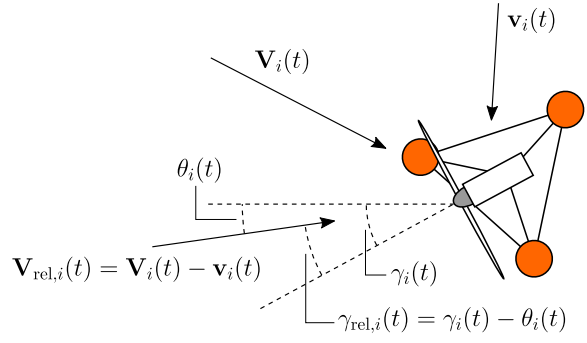}
	\caption{Schematic of floating platform velocity vector $\mathbf{v}_i(t)$, incident wind velocity vector $\mathbf{V}_i(t)$, and the relative incident velocity vector $\mathbf{V}_{\mathrm{rel},i}(t)$ at the location of turbine~$i$.}
	\label{Figure - Effective incident wind speed}
\end{figure}

The total power output of the wind farm $P_\mathrm{farm}(t)$ is computed as the sum of electricity production from all wind turbines as follows:
\begin{equation}
	P_\mathrm{farm}(t) = \sum_{i \in \mathscr{F}} P_i(t),
\end{equation}
where $P_i(t)$ is the power output of turbine~$i$, and is estimated assuming steady-state performance as follows~\cite{Manwell2009}:
\begin{equation} \label{Equation - Turbine power output}
	P_i(t) = \frac{1}{8} C_{\mathrm{p},i}(t) \rho_\mathrm{a} \pi D_i^2 \left\Vert \mathbf{V}_{\mathrm{rel},i}(t) \right\Vert^3.
\end{equation}
$D_i$ is rotor diameter of turbine~$i$, $\rho_\mathrm{a}$ is the density of air, and $\mathbf{V}_{\mathrm{rel},i}(t)$ is the wind velocity that is incident upon the rotor of turbine~$i$ from the perspective of an observer who is fixed to turbine~$i$. Referring to Fig.~\ref{Figure - Effective incident wind speed}, $\mathbf{V}_{\mathrm{rel},i}(t)$ is defined as follows:
\begin{equation}
	\mathbf{V}_{\mathrm{rel},i}(t) = \mathbf{V}_i(t) - \mathbf{v}_i(t),
\end{equation}
where $\mathbf{v}_i(t)$ is the velocity vector of turbine~$i$, and $\mathbf{V}_i(t)$ is the wind velocity vector (in the global frame) that is incident upon the rotor of turbine~$i$ with the following $\hat{x}$ and $\hat{y}$ components:
\begin{equation}
	\mathbf{V}_i(t) \coloneqq
	\begin{bmatrix}
		U_i(t) & V_i(t)
	\end{bmatrix}^\mathrm{T}.
\end{equation}
$\mathbf{V}_i(t)$ is calculated using the wake interaction model discussed in Section~\ref{Subsection - Wind interaction model}.

The power coefficient $C_{\mathrm{p},i}(t)$ of turbine~$i$ is computed based on the vortex cylinder model of a yawed actuator disc as follows~\cite{Burton2011}:
\begin{equation} \label{Equation - Power coefficient}
	C_{\mathrm{p},i}(t) = 4 a_i(t) \left( \cos \gamma_{\mathrm{rel},i}(t) - a_i(t) \right) \Bigg( \cos \gamma_{\mathrm{rel},i}(t) + \tan \frac{\chi_i(t)}{2} \sin \gamma_{\mathrm{rel},i}(t) - a_i(t) \sec^2 \frac{\chi_i(t)}{2} \Bigg),
\end{equation}
where $a_i(t)$ is the axial induction factor of turbine~$i$ and, as per Fig.~\ref{Figure - Effective incident wind speed}, $\gamma_{\mathrm{rel},i}(t)$ is the yaw misalignment of turbine~$i$ relative to $\mathbf{V}_{\mathrm{rel},i}(t)$ as follows:
\begin{equation}
	\gamma_{\mathrm{rel},i}(t) = \gamma_i(t) - \theta_i(t).
\end{equation}
In the above expression, $\gamma_i(t)$ is the yaw angle of turbine~$i$ and $\theta_i(t)$ is the angle of $\mathbf{V}_{\mathrm{rel},i}(t)$ relative to the positive $\hat{x}$ axis as follows:
\begin{equation}
	\theta_i(t) = \tan^{-1} \frac{V_i(t) - v_{y,i}(t)}{U_i(t) - v_{x,i}(t)}.
\end{equation}
Finally, $\chi_i(t)$ is the wake skew angle immediately past the rotor of turbine~$i$ and is approximated as follows~\cite{Burton2011}:
\begin{equation}
	\chi_i(t) = \left( 0.6 a_i(t) + 1 \right) \gamma_{\mathrm{rel},i}(t).
\end{equation}

\subsection{Floating wind turbine motion}

The rates of change of the position and velocity of turbine~$i$ are expressed as follows:
\begin{eqnarray}
	\dot{\mathbf{r}}_i(t) & = & \mathbf{v}_i(t), \\
	\dot{\mathbf{v}}_i(t) & = & \frac{\mathbf{F}_i(t)}{m_i + m_{\mathrm{a},i}},
\end{eqnarray}
where $m_i$ is the mass of floating wind turbine~$i$. The added mass\footnote{Added mass accounts for hydrodynamic loads that act upon an object that is accelerating with respect to the surrounding fluid. It compounds with hydrodynamic drag forces, which are typically modeled as functions of instantaneous velocity only.} $m_{\mathrm{a},i}$ associated with turbine~$i$ will be discussed along with the hydrodynamic drag force.

\begin{figure}
	\centering
	\includegraphics[width=2.5in]{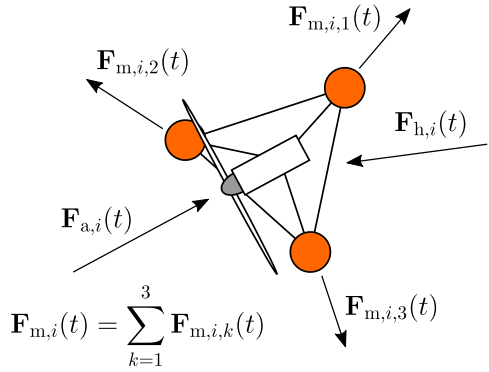}
	\caption{Schematic of aerodynamic thrust force $\mathbf{F}_{\mathrm{a},i}(t)$, hydrodynamic drag force $\mathbf{F}_{\mathrm{h},i}(t)$, and mooring line forces $\mathbf{F}_{\mathrm{m},i,k}(t)$ acting on wind turbine~$i$ with a semi-submersible floating platform.}
	\label{Figure - Turbine forces}
\end{figure}

As shown in Fig.~\ref{Figure - Turbine forces}, the total force $\mathbf{F}_i(t)$ acting on turbine~$i$ is the sum of its respective aerodynamic, hydrodynamic, and mooring line forces as follows:
\begin{equation}
	\mathbf{F}_i(t) = \mathbf{F}_{\mathrm{a},i}(t) + \mathbf{F}_{\mathrm{h},i}(t) + \mathbf{F}_{\mathrm{m},i}(t).
\end{equation}

The aerodynamic thrust force $\mathbf{F}_{\mathrm{a},i}(t)$ acting on the rotor of turbine~$i$ is expressed as follows:
\begin{equation} \label{Equation - Thrust force}
	\mathbf{F}_{\mathrm{a},i}(t) = \frac{1}{8} C_{\mathrm{t},i}(t) \rho_\mathrm{a} \pi D_i^2 \left\Vert \mathbf{V}_{\mathrm{rel},i}(t) \right\Vert^2 \mathbf{n}_i(t),
\end{equation}
where the thrust coefficient $C_{\mathrm{t},i}(t)$ is computed based on the vortex cylinder model of a yawed actuator disc as follows~\cite{Burton2011}:
\begin{equation} \label{Equation - Thrust coefficient}
	C_{\mathrm{t},i}(t) = 4 a_i(t) \left( \cos \gamma_{\mathrm{rel},i}(t) + \tan \frac{\chi_i(t)}{2} \sin \gamma_{\mathrm{rel},i}(t) - a_i(t) \sec^2 \frac{\chi_i(t)}{2} \right),
\end{equation}
and $\mathbf{n}_i(t)$ is a unit vector normal to the rotor of turbine~$i$ as follows:
\begin{equation}
	\mathbf{n}_i(t) =
	\begin{bmatrix}
		\cos \gamma_i(t) & \sin \gamma_i(t)
	\end{bmatrix}^\mathrm{T}.
\end{equation}

Based on elementary fluid mechanics principles concerning immersed bodies, $\mathbf{F}_{\mathrm{h},i}(t)$ is approximated by summing the drag force contributions of all submerged components of turbine~$i$ as follows:
\begin{equation} \label{Equation - Drag force}
	\mathbf{F}_{\mathrm{h},i}(t) = \frac{1}{2} \left( \sum_{j \in \mathscr{D}_i} C_{\mathrm{d},i,j} A_{\mathrm{d},i,j} \right) \rho_\mathrm{w} \left\Vert \mathbf{w}(t) - \mathbf{v}_i(t) \right\Vert \left( \mathbf{w}(t) - \mathbf{v}_i(t) \right),
\end{equation}
where $\rho_\mathrm{w}$ is the density of ocean water, and $\mathbf{w}(t)$ is the ocean current velocity vector (which we assume to be $\mathbf{w}(t) = 0\,\si{m/s}$ in this work). Let the set $\mathscr{D}_i = \left\{ 1, 2, \cdots, N_{\mathrm{h},i} \right\}$ denote the indices of all submerged components that contribute to the hydrodynamic drag force acting on turbine~$i$, with $N_{\mathrm{h},i}$ being equal to the total number of submerged components of turbine~$i$. $C_{\mathrm{d},i,j}$ and $A_{\mathrm{d},i,j}$ are thereby the drag coefficient and reference area of the $j^\mathrm{th}$ submerged component of turbine~$i$.

In a similar manner, the total added mass $m_{\mathrm{a},i}$ associated with turbine~$i$ is estimated by summing the added mass contributions of all submerged components of turbine~$i$ as follows:
\begin{eqnarray}
	\label{Equation - Added mass}
	m_{\mathrm{a},i} = \rho_\mathrm{w} \sum_{j \in \mathscr{D}_i} C_{\mathrm{a},i,j} A_{\mathrm{a},i,j},
\end{eqnarray}
where $C_{\mathrm{a},i,j}$ is the added mass coefficient of the $j^\mathrm{th}$ submerged component of turbine~$i$, and $A_{\mathrm{a},i,j}$ is the added mass reference area of the same component. 

Let the set $\mathscr{M}_i = \left\{ 1, 2, \cdots, N_{\mathrm{m},i} \right\}$ denote the indices of all mooring lines connected to turbine~$i$, with $N_{\mathrm{m},i}$ being equal to the total number of mooring lines attached to turbine~$i$. $\mathbf{F}_{\mathrm{m},i}(t)$ may then be expressed as the sum of all mooring force contributions acting on turbine~$i$ as follows:
\begin{equation}
	\mathbf{F}_{\mathrm{m},i}(t) = \sum_{k \in \mathscr{M}_i} \mathbf{F}_{\mathrm{m},i,k}(t),
\end{equation}
where $\mathbf{F}_{\mathrm{m},i,k}(t)$ is the restoring force exerted on turbine~$i$ by its $k^\mathrm{th}$ mooring line. This force is calculated by first finding the magnitude of the horizontal component of tension within mooring line $k$ of turbine~$i$, and then projecting this tension in the appropriate direction as follows:
\begin{equation} \label{Equation - Mooring force k}
	\mathbf{F}_{\mathrm{m},i,k}(t) = - H_{\mathrm{F},i,k}(t) \frac{\mathbf{r}_{\mathrm{F}/\mathrm{A},i,k}(t)}{\left\Vert \mathbf{r}_{\mathrm{F}/\mathrm{A},i,k}(t) \right\Vert}.
\end{equation}
The function $H_{\mathrm{F},i,k}(t)$ outputs the horizontal component of tension along the $k^\mathrm{th}$ mooring line of turbine~$i$. This function is generated by solving the static differential equations describing a suspended cable which is either partially contacting or fully lifted above the seabed. The relevant formulae are provided in Appendix~\ref{Appendix - Catenary solution} and are also available in out previous publication~\cite{Kheirabadi2020}.

\begin{figure}
	\centering
	\includegraphics[width=3in]{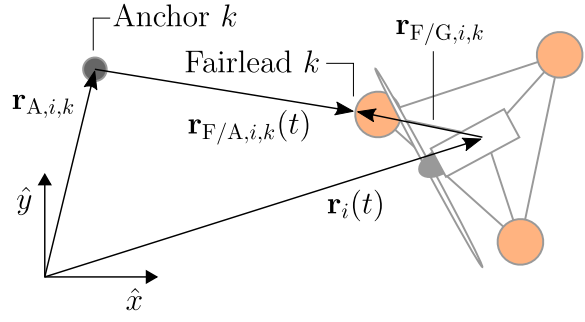}
	\caption{Schematic of position vectors that are relevant for calculating the force in mooring line $k$ of turbine~$i$.}
	\label{Figure - Turbine position vectors}
\end{figure}

As shown in Fig.~\ref{Figure - Turbine position vectors}, the term $\mathbf{r}_{\mathrm{F}/\mathrm{A},i,k}(t)$ describes the position vector from the anchor of the $k^\mathrm{th}$ mooring line of turbine~$i$ to the corresponding fairlead, and is expressed as follows:
\begin{equation}
	\mathbf{r}_{\mathrm{F}/\mathrm{A},i,k}(t) = \mathbf{r}_i(t) + \mathbf{r}_{\mathrm{F}/\mathrm{G},i,k} - \mathbf{r}_{\mathrm{A},i,k},
\end{equation}
where $\mathbf{r}_{\mathrm{F}/\mathrm{G},i,k}$ is a constant position vector from the center-of-gravity of turbine~$i$ to the fairlead that connects to the $k^\mathrm{th}$ mooring line of the same turbine, and $\mathbf{r}_{\mathrm{A},i,k}$ is a constant position vector representing the location of the anchor of the same mooring line. In Eq.~(\ref{Equation - Mooring force k}), dividing $\mathbf{r}_{\mathrm{F}/\mathrm{A},i,k}(t)$ by its Euclidean norm therefore produces a unit vector that points from the anchor of the $k^\mathrm{th}$ mooring line of turbine~$i$ to the corresponding fairlead. The restoring force associated with this mooring line pulls the turbine in the opposite direction.

\subsection{Single wake model} \label{Subsection - Single wake model}

\begin{figure}
	\centering
	\includegraphics[width=4in]{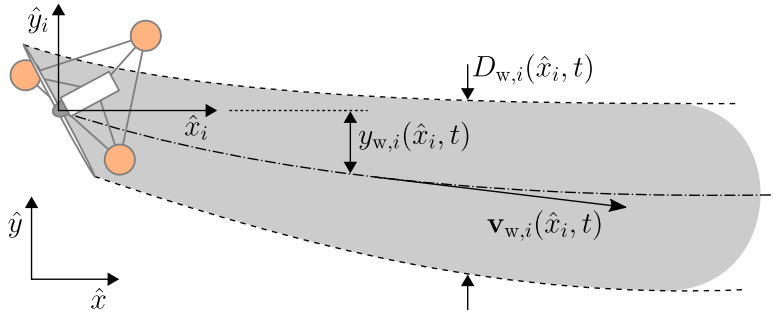}
	\caption{Schematic of characteristics necessary for modeling the wake generated by turbine~$i$. The wake centerline position $y_{\mathrm{w},i}(\hat{x}_i,t)$, average wake velocity $\mathbf{v}_{\mathrm{w},i}(\hat{x}_i,t)$, and wake diameter $D_{\mathrm{w},i}(\hat{x}_i,t)$ are defined within the reference frame that is fixed to turbine~$i$.}
	\label{Figure - Wake characteristics}
\end{figure}

Fig.~\ref{Figure - Wake characteristics} shows the characteristics of interest when modeling wake~$i$, which is the wake generated by the rotor of turbine~$i$. These characteristics include the wake's centerline position $y_{\mathrm{w},i}(\hat{x}_i,t)$ relative to the $\hat{x}_i$ axis, its average velocity vector $\mathbf{v}_{\mathrm{w},i}(\hat{x}_i,t)$ measured in frame~$i$, and its diameter $D_{\mathrm{w},i}(\hat{x}_i,t)$.

Two key assumptions are necessary for justifying the mathematical formulation presented in this section. First, if fluctuations in the wind direction relative to the $\hat{x}_i$ axes are presumed to be small, then all wake characteristics may be defined as smooth functions of only $\hat{x}_i$ and $t$. Furthermore, wake cross-sections may be assumed to always remain normal to the predominant flow direction, which corresponds to the positive $\hat{x}$ and $\hat{x}_i$ axes in our work.

Second, if the free stream wind speed is presumed to be significantly larger than the velocities of floating platforms, then the equations of motion describing any wake may be defined relative to a reference frame that is fixed to the wake-generating turbine. The frame of reference shown in Fig.~\ref{Figure - Wake characteristics} is therefore non-inertial and translates with turbine~$i$, while $y_{\mathrm{w},i}(\hat{x}_i,t)$, $\mathbf{v}_{\mathrm{w},i}(\hat{x}_i,t)$, and $D_{\mathrm{w},i}(\hat{x}_i,t)$ are defined in this translating frame. This approach eliminates the need to model wake behavior upstream of turbine~$i$, while removing time-dependency from the wake centerline boundary condition (\textit{i.e.} $y_{\mathrm{w},i}(\hat{x}_i,t)$ is always equal to zero at $\hat{x}_i = 0\,\si{m}$). 

Granting these preliminaries, the equations of motion describing wake~$i$ may now be derived. Specifically, we shall present partial differential equations that model wake average velocities, wake centerline locations, and wake diameters over space and time. Let the vector $\mathbf{L}_i(\hat{x}_i,t)$ describe the linear momentum deficit of wake~$i$ per unit length along the $\hat{x}_i$ axis as follows:
\begin{equation}
	\mathbf{L}_i(\hat{x}_i,t) = \rho_a \frac{\pi}{4} D_{\mathrm{w},i}^2(\hat{x}_i,t) \left[ \mathbf{V}_\infty(t) - \left( \mathbf{v}_i(t) + \mathbf{v}_{\mathrm{w},i}(\hat{x}_i,t) \right) \right].
\end{equation}
As $\mathbf{v}_{\mathrm{w},i}(\hat{x}_i,t)$ is measured in frame~$i$, the term $\mathbf{v}_i(t) + \mathbf{v}_{\mathrm{w},i}(\hat{x}_i,t)$ redefines the velocity of wake~$i$ in the global frame. Since no external forces impact wake~$i$, the time-derivative of $\mathbf{L}_i(\hat{x}_i,t)$ must equate to zero, which results in the following momentum conservation equation:
\begin{multline} \label{Equation - Momentum conservation}
	\frac{\partial \mathbf{v}_{\mathrm{w},i}(\hat{x}_i,t)}{\partial t} + \left( U_\infty(t) - v_{x,i}(t) \right) \frac{\partial \mathbf{v}_{\mathrm{w},i}(\hat{x}_i,t)}{\partial \hat{x}_i} =\\
	\dot{\mathbf{V}}_\infty(t) - \dot{\mathbf{v}}_i(t) + \frac{2}{D_{\mathrm{w},i}(\hat{x}_i,t)} \frac{d D_{\mathrm{w},i}(\hat{x}_i,t)}{d t} \left( \mathbf{V}_\infty(t) - \mathbf{v}_i(t) - \mathbf{v}_{\mathrm{w},i}(\hat{x}_i,t) \right).
\end{multline}
The time-derivative of $y_{\mathrm{w},i}(\hat{x}_i,t)$ must equate to the $\hat{y}_i$ component of $\mathbf{v}_{\mathrm{w},i}(\hat{x}_i,t)$, which results in the following expression describing the wake centerline location:
\begin{equation} \label{Equation - Wake centerline ROT}
	\frac{\partial y_{\mathrm{w},i}(\hat{x}_i,t)}{\partial t} + \left( U_\infty(t) - v_{x,i}(t) \right) \frac{\partial y_{\mathrm{w},i}(\hat{x}_i,t)}{\partial \hat{x}_i} = v_{\mathrm{w},i}(\hat{x}_i,t).
\end{equation}
In Eqs.~(\ref{Equation - Momentum conservation}) and~(\ref{Equation - Wake centerline ROT}), $u_{\mathrm{w},i}(\hat{x}_i,t)$ and $v_{\mathrm{w},i}(\hat{x}_i,t)$ are the $\hat{x}_i$ and $\hat{y}_i$ components of $\mathbf{v}_{\mathrm{w},i}(\hat{x}_i,t)$, $v_{x,i}(t)$ is the velocity of turbine~$i$ in the $\hat{x}$ direction, $U_\infty(t)$ is the free stream wind speed in the $\hat{x}$ direction, and the term $U_\infty(t) - v_{x,i}(t)$ serves as the transport speed in the $\hat{x}_i$ direction. When modeling fluids using the three-dimensional Navier-Stokes equations, the transport and fluid velocities at any given point are equal. Following this logic, the transport speed in Eqs.~(\ref{Equation - Momentum conservation}) and~(\ref{Equation - Wake centerline ROT}) should simply be $u_{\mathrm{w},i}(\hat{x}_i,t)$. However, when neglecting three-dimensional effects, it is debatable exactly how the transport velocity should be defined. Our simulations indicate that defining the transport speed as the free stream wind speed (defined in frame~$i$) yields predictions closer to experimental observations than does setting the transport speed to $u_{\mathrm{w},i}(\hat{x}_i,t)$.

In steady-state parametric wake models, the wake diameter is typically assumed to grow at a constant spatial expansion rate $k_x$ along the downstream direction. When modeling wakes dynamically, however, we assume that wake diameters grow at a constant temporal expansion rate $k_t$. In other words, the time-derivative of $D_{\mathrm{w},i}(\hat{x}_i,t)$ must equate to $k_t$ as follows:
\begin{equation} \label{Equation - Wake diameter}
	\frac{\partial D_{\mathrm{w},i}(\hat{x}_i,t)}{\partial t} + \left( U_\infty(t) - v_{x,i}(t) \right) \frac{\partial D_{\mathrm{w},i}(\hat{x}_i,t)}{\partial \hat{x}_i} = k_t.
\end{equation}
If the spatial expansion rate $k_x$ under steady-state conditions is known for some reference free stream wind speed $U_{\infty,\mathrm{ref}}$, the temporal expansion rate at $U_{\infty,\mathrm{ref}}$ must be $k_t = k_x U_{\infty,\mathrm{ref}}$. Assuming that the free stream wind speed $\left\Vert \mathbf{V}_\infty(t) \right\Vert$ does not vary significantly from $U_{\infty,\mathrm{ref}}$, then $k_t$ may be assumed to remain constant.

In order to obtain the wake states employed in Eq.~(\ref{Equation - Wake point state vector}), the spatial gradients in Eqs.~(\ref{Equation - Momentum conservation}), (\ref{Equation - Wake centerline ROT}), and (\ref{Equation - Wake diameter}) must be discretized over some fixed downstream distance using the finite difference method, which would yield a system of nonlinear ordinary differential equations that would be rearranged to state-space form. We will not present the discretized forms of these equations as the finite difference method is an elementary numerical technique.

When implementing the above solution, we recommend the following initial conditions:
\begin{eqnarray}
	\label{Equation - Initial cond. wake centerline}
	y_{\mathrm{w},i}(\hat{x}_i,0) & = & \frac{V_\infty(0)}{U_\infty(0)} \hat{x}_i, \\
	\label{Equation - Initial cond. wake velocity}
	\mathbf{v}_{\mathrm{w},i}(\hat{x}_i,0) & = & \mathbf{V}_\infty(0) - \mathbf{v}_i(0), \\
	\label{Equation - Initial cond. wake diameter}
	D_{\mathrm{w},i}(\hat{x}_i,0) & = & D_i + k_x \hat{x}_i,
\end{eqnarray}
which ensure, respectively, that all wake centerlines are initially aligned with the free stream wind, wake velocities are initially equal to the free stream wind velocity, and that wake diameters initially grow at a predefined spatial rate $k_x$. Note that $D_i$ is the diameter of turbine~$i$. With regards to boundary conditions, the following are necessary based on assumptions inherent to FOWFSim-Dyn:
\begin{eqnarray}
	\label{Equation - Centerline boundary condition}
	y_{\mathrm{w},i}(0,t) & = & 0, \\
	\label{Equation - Velocity boundary condition}
	\mathbf{v}_{\mathrm{w},i}(0,t) & = & \mathbf{v}_{\mathrm{w,init},i}(t),\\
	\label{Equation - Diameter boundary condition}
	D_{\mathrm{w},i}(0,t) & = & D_i,
\end{eqnarray}
Equation~(\ref{Equation - Centerline boundary condition}) states that the centerline of wake~$i$ at $\hat{x}_i = 0\,\si{m}$ must always correspond to the location of turbine~$i$, which is in fact the origin of frame~$i$. Equation~(\ref{Equation - Velocity boundary condition}) states that the velocity of wake~$i$ at $\hat{x}_i = 0\,\si{m}$ must always be equal to the wake velocity $\mathbf{v}_{\mathrm{w,init},i}(t)$ immediately downstream of the rotor of turbine~$i$. Finally, Eq.~(\ref{Equation - Diameter boundary condition}) requires that the diameter of wake~$i$ at the location of turbine~$i$ is always equal to the rotor diameter of this turbine.

We calculate the velocity vector $\mathbf{v}_{\mathrm{w,init},i}(t)$ based on simplifications made to Glauert's momentum theory~\cite{Burton2011} by Bastankhah and Port\'e Agel~\cite{Bastankhah2016} as follows:
\begin{equation} \label{Equation - Initial wake velocity}
	\mathbf{v}_{\mathrm{w,init},i}(t) = \left\Vert \mathbf{V}_{\mathrm{rel},i}(t) \right\Vert \sqrt{1 - C_{\mathrm{t},i}(t)}
	\begin{bmatrix}
		\cos \left( \xi_{\mathrm{w,init},i}(t) + \theta_i(t) \right)\\
		\sin \left( \xi_{\mathrm{w,init},i}(t) + \theta_i(t) \right)
	\end{bmatrix},
\end{equation}
where $\xi_{\mathrm{w,init},i}(t)$ is the initial wake skew angle, which is expressed as follows based on a momentum conservation derivation reported by Jim\'enez \textit{et al.}~\cite{Jimenez2009}:
\begin{equation}
	\xi_{\mathrm{w,init},i}(t) = - \frac{C_{\mathrm{t},i}(t)}{2} \cos^2 \gamma_{\mathrm{rel},i}(t) \sin \gamma_{\mathrm{rel},i}(t).
\end{equation}
The derivation by Bastankhah and Port\'e Agel~\cite{Bastankhah2016} assumes that the free stream wind velocity is aligned with the $\hat{x}$ axis. As a result, the addition of $\theta_i(t)$ to $\xi_{\mathrm{w,init},i}(t)$ in Eq.~(\ref{Equation - Initial wake velocity}) accounts for the misalignment of $\mathbf{V}_\infty(t)$ relative to the $\hat{x}$ axis.

\subsection{Wake interaction model} \label{Subsection - Wind interaction model}

When a wind turbine rotor is influenced by wakes that are generated from multiple upstream turbines, a wake interaction model is necessary for approximating the resultant effective wind speed that is incident on the downstream rotor. The most commonly used wake interaction technique is based on the assumption that the effective kinetic energy deficit at the location of the downstream rotor must be equal to the sum of kinetic energy deficits of all pertinent wakes~\cite{Katic1986}. As a result, the effective wind speed at the downstream rotor is a function of the root-sum-square of relevant wake velocity deficits. Further enhancement may be obtained by approximating wake velocity profiles using Gaussian distributions~\cite{Bastankhah2016}. We continue to make use of this wake interaction methodology.

Let the set $\mathscr{U}_i = \left\{ 1, 2, \cdots, i - 1 \right\}$ denote the indices of all turbines that are located upstream of turbine~$i$. The effective wind velocity vector that is incident on the rotor of turbine~$i$ may therefore be expressed as follows:
\begin{equation} \label{Equation - Wake interaction model}
	\mathbf{V}_i(t) = \left\{ \left\Vert \mathbf{V}_\infty(t) \right\Vert - \sqrt{\sum_{q \in \mathscr{U}_i} \left( \left\Vert \mathbf{V}_\infty(t) \right\Vert - \overline{\mathbf{v}}_{\mathrm{w},q \rightarrow i}(t) \cdot \mathbf{n}_\infty(t) \right)^2} \right\} \mathbf{n}_\infty(t).
\end{equation}
where $\mathbf{n}_\infty(t)$ is a unit vector aligned with $\mathbf{V}_\infty(t)$ as follows:
\begin{equation}
	\mathbf{n}_\infty(t) = \frac{\mathbf{V}_\infty(t)}{\left\Vert \mathbf{V}_\infty(t) \right\Vert},
\end{equation}
and $\overline{\mathbf{v}}_{\mathrm{w},q \rightarrow i}(t)$ is the \textit{effective} velocity of wake~$q$ that is incident upon the rotor of wake~$i$. Equation~(\ref{Equation - Wake interaction model}) projects $\overline{\mathbf{v}}_{\mathrm{w},q \rightarrow i}(t)$ along the free stream wind direction (hence the dot product operation with $\mathbf{n}_\infty(t)$), and then computes the velocity deficit in this direction. Average wake velocities perpendicular to the free stream wind direction are assumed to be negligibly small far enough downstream; their effects are therefore neglected.

We now describe our procedure for computing $\overline{\mathbf{v}}_{\mathrm{w},q \rightarrow i}(t)$. Let $\mathbf{v}_{\mathrm{w},q \rightarrow i}(t)$ denote the average velocity of wake~$q$ at the location of wake~$i$ as follows:
\begin{equation}
	\label{Equation - Project wind velocity}
	\mathbf{v}_{\mathrm{w},q \rightarrow i}(t) = \mathbf{v}_q(t) + \mathbf{v}_{\mathrm{w},q}(x_i(t) - x_q(t), t).
\end{equation}
Since the average velocity vector of wake~$q$ is defined in frame~$q$, the substitution $\hat{x}_q = x_i(t) - x_q(t)$ into $\mathbf{v}_{\mathrm{w},q}(\hat{x}_q,t)$ is necessary for identifying the location of turbine~$i$ in frame~$q$. The addition of the turbine velocity vector $\mathbf{v}_q(t)$ then transforms $\mathbf{v}_{\mathrm{w},q}(x_i(t) - x_q(t), t)$ to the global frame.

The next step is to generate a Gaussian profile $\breve{\mathbf{v}}_{\mathrm{w},q \rightarrow i}(r,t)$, where $r$ is the radial distance from the centerline of wake~$q$, to approximate the continuous velocity distribution of wake~$q$ at the location of wake~$i$. Imposing a requirement that the total momentum deficit of $\mathbf{V}_\infty(t) - \breve{\mathbf{v}}_{\mathrm{w},q \rightarrow i}(r,t)$ per unit length must equate that of a top-hat distribution with amplitude $\mathbf{V}_\infty(t) - \mathbf{v}_{\mathrm{w},q \rightarrow i}(t)$ as follows:
\begin{equation}
	\int_{0}^{\infty} \rho_a 2 \pi r \left( \mathbf{V}_\infty(t) - \breve{\mathbf{v}}_{\mathrm{w},q \rightarrow i}(r,t) \right) d r = \rho_a \frac{\pi}{4} D_{\mathrm{w},q \rightarrow i}(t) \left( \mathbf{V}_\infty(t) - \mathbf{v}_{\mathrm{w},q \rightarrow i}(t) \right),
\end{equation}
the following Gaussian profile is then obtained:
\begin{equation} \label{Equation - Gaussian profile}
	\mathbf{V}_\infty(t) - \breve{\mathbf{v}}_{\mathrm{w},q \rightarrow i}(r,t) = \frac{1}{8} \left( \frac{D_{\mathrm{w},q \rightarrow i}(t)}{\sigma} \right)^2 \left( \mathbf{V}_\infty(t) - \mathbf{v}_{\mathrm{w},q \rightarrow i}(t) \right) \exp{\frac{-r^2}{2 \sigma^2}},
\end{equation}
where $D_{\mathrm{w},q \rightarrow i}(t)$ is the diameter of wake~$q$ at the location of turbine~$i$ as follows:
\begin{equation}
	D_{\mathrm{w},q \rightarrow i} = D_{\mathrm{w},q}(x_i(t) - x_q(t), t).
\end{equation}
The standard deviation $\sigma$ in Eq.~(\ref{Equation - Gaussian profile}) may be estimated based on experimental or high-fidelity numerical data.

Finally, the effective velocity $\overline{\mathbf{v}}_{\mathrm{w},q \rightarrow i}(t)$ is obtained by averaging $\breve{\mathbf{v}}_{\mathrm{w},q \rightarrow i}(r,t)$ along the rotor area $A_i$ of turbine~$i$. This task is achieved by numerically computing the following integral at each time-step:
\begin{equation}
	\overline{\mathbf{v}}_{\mathrm{w},q \rightarrow i}(t) = \frac{4}{\pi D_i^2} \int_{A_i} \breve{\mathbf{v}}_{\mathrm{w},q \rightarrow i}(r,t) d A.
\end{equation}

\subsection{Model limitations} \label{Subsection - Assumptions and limitations}

Several assumptions have been made when developing FOWFSim-Dyn which impose limitations on its fidelity and applicability. The current subsection summarizes these limitations.

\subsubsection{Two-dimensional floating wind turbine dynamics}

The first and most crucial of these assumptions is that floating platform motion may be adequately captured using a two-dimensional planar model. That is to say, we neglect floating platform heave, yaw, pitch, and roll. In consequence, FOWFSim-Dyn fails to capture dynamic effects induced by ocean waves and oscillatory wind conditions on platform rotation. FOWFSim-Dyn remains appropriate for wind farm controller design and testing, since this application is primarily concerned with average rotor positions over extended periods of time. However, any attempt to control or evaluate individual wind turbine dynamics requires the use of three-dimensional multi-body nonlinear modeling tools.

\subsubsection{Steady-state mooring line model}

Although we present a dynamic model, mooring line tensions are found based on the solution to a static suspended cable problem. It has been reported by Hall \textit{et al.}~\cite{Hall2014} that such static models accurately predict mooring line loads and floating wind turbine motion; thus rendering them appropriate for wind farm control. However, Hall \textit{et al.}~\cite{Hall2014} also mentioned that use of such models may lead to large inaccuracies in turbine load predictions. Therefore, analysis and control of individual turbine motion must consider higher-fidelity modeling techniques such as a lumped-mass dynamic mooring line model~\cite{Hall2015}.

\subsubsection{Steady-state turbine aerodynamics}

Turbine power outputs and thrust forces (Eqs.~(\ref{Equation - Turbine power output}) and (\ref{Equation - Thrust force})), along with their respective coefficients (Eqs.~(\ref{Equation - Power coefficient}) and (\ref{Equation - Thrust coefficient})), are calculated based on steady-state actuator disc theory. This approach assumes ideal rotors and fails to capture unsteady aerodynamic effects and asymmetric rotor loadings. These phenomena significantly influence blade loads when yaw misalignment occurs; however, for the purpose of wind farm control, our focus lies on the overall influence of rotor operation on fully-developed wake regions. Nonetheless, any turbine-level analysis requires more detailed fluid-structure interaction modeling.

The computation of $\mathbf{v}_{\mathrm{w,init},i}(t)$ in Eq.~(\ref{Equation - Initial wake velocity}), which is the average wake velocity immediately downstream of turbine~$i$, relies on a steady-state momentum balance on a control volume spanning across the rotor of turbine~$i$. As a result, momentum fluxes into and out of this control volume are considered, while the rate-of-change of momentum within the control volume is neglected. Given the low density of air, these inertial effects may be neglected, although their significance should be investigated.

\subsubsection{Sources of wake deflection}

FOWFSim-Dyn does not capture wake centerline deflection caused by rotor rotation. This phenomenon was first observed in high-fidelity simulations conducted by Gebraad \textit{et al.}~\cite{Gebraad2016}; however, more recent work by Fleming \textit{et al.}~\cite{Fleming2018} showed that the scale of this phenomenon is insignificant. Instead, Fleming \textit{et al.}~\cite{Fleming2018} observed that vortices generated by turbine rotors induce wake deflection past downstream machines, even if their rotors are not operated with yaw offset. Additional terms may be added to Eqs.~(\ref{Equation - Momentum conservation}) and (\ref{Equation - Wake centerline ROT}) to account for such phenomena.

\subsubsection{Spatial-uniformity and consistency of the free stream wind}

In the current paper, we have assumed that the free stream wind velocity is uniform throughout the wind farm, which is why the variable $\mathbf{V}_\infty(t)$ is solely a function of time. This variable may readily be expressed as $\mathbf{V}_\infty(\hat{x},t)$ if spatial variations of the free stream wind velocity are known. Furthermore, in order to represent wake characteristics purely as a function of the downstream distance along the $\hat{x}_i$ axes, while ignoring changes in the cross-sectional areas of wakes, we assumed that variations in the free stream wind direction are small relative to the $\hat{x}$ axis. The $\hat{y}$ component of the free stream wind velocity must therefore remain small in comparison to its $\hat{x}$ component.

\section{Simulation results and discussions} \label{Section - Model tuning and validation}

In this section, we first perform a mesh sensitivity analysis to ascertain the dependency of model predictions upon the size of finite difference elements in Section~\ref{Subsection - Mesh analysis}. We then validate FOWFSim-Dyn against steady-state experimental results reported by Bastankhah and Port\'e Agel~\cite{Bastankhah2016} in Section~\ref{Subsection - Validation}. Finally, we present dynamic simulation results for various scenarios to ensure that model predictions are in line with physical expectations and intuition in Section~\ref{Subsection - Dynamic simulation}.

\subsection{Mesh sensitivity analysis} \label{Subsection - Mesh analysis}

For a mesh sensitivity study, we simulate the experimental setup employed by Bastankhah and Port\'e Agel~\cite{Bastankhah2016}. Namely, the wake of a single fixed-foundation turbine with diameter $D = 15\,\si{cm}$ is simulated with a steady free stream wind speed of $U_\infty = 4.88\,\si{m/s}$. The turbine's axial induction factor is set to the optimal value of $a = 1/3$ and a yaw angle of $\gamma = 20\,\si{deg}$ is implemented to observe mesh effects on wake deflection. To approximate steady-state results, all simulations are run for a duration of $5\,\si{sec}$ and data is extracted from the final time-step. The Gaussian profile standard deviation is set to $\sigma = 0.025 \hat{x} + 0.396\,\si{m}$ based on experimental data\footnote{Velocity profiles corresponding to a yaw angle of $\gamma = 0\,\si{deg}$ from Fig. 21 in the paper by Bastankhah and Port\'e Agel~\cite{Bastankhah2016} were digitized and Gaussian function curve fitting was used to compute the standard deviation.} reported by Bastankhah and Port\'e Agel~\cite{Bastankhah2016}. The (diametrical) spatial wake expansion constant is set to $k_x = 0.08$ as per the recommendation by Shakoor \textit{et al.}~\cite{Shakoor2016}.

\begin{figure}
	\centering
	\includegraphics[width=4.5in]{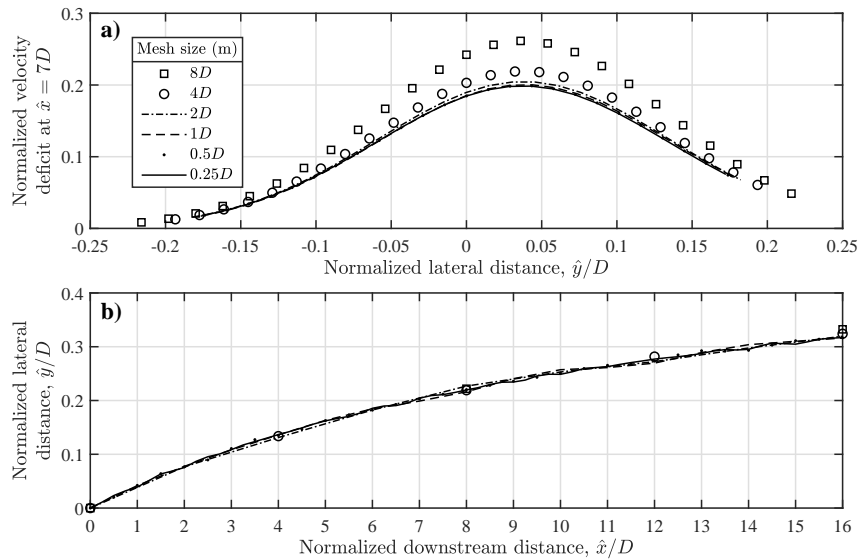}
	\caption{Effects of various finite difference mesh element sizes on \textbf{a)} the steady-state wake velocity profile at a downstream distance of $7 D$, and \textbf{b)} the steady-state wake centerline evolution. Simulation parameters: $D = 15\,\si{cm}$, $U_\infty = 8\,\si{m/s}$, $a = 1/3$, $\gamma = 20\,\si{deg}$, $k_x = 0.08$, $\sigma = 0.025 \hat{x} + 0.396\,\si{m}$.}
	\label{Figure - Mesh convergence results}
\end{figure}

Simulated wake centerlines and normalized velocity deficit profiles at a downstream distance of $7 D$ are plotted in Fig.~\ref{Figure - Mesh convergence results} for different finite difference element sizes. Qualitatively, it is apparent that the evolution of the wake centerline is insignificantly influenced by the mesh size. At $\hat{x}/D = 16$, the centerline deflection obtained using an element size of $8 D$ only differs by $5\,\si{\percent}$ relative to the value corresponding to an element size of $0.25 D$. As a result, we solely utilize the maximum normalized velocity deficit as a convergence criterion.

Fig.~\ref{Table - Mesh convergence results} lists the computation times corresponding to different element sizes from Fig.~\ref{Figure - Mesh convergence results} as well as predicted maximum normalized velocity deficits. Dynamic simulations were performed using the MATLAB fourth-order Runge-Kutta solver implemented on a laptop computer with a $2.80\,\si{GHz}$ Intel Core i7-7700HQ processor. Figure~\ref{Table - Mesh convergence results} also lists the convergence of the maximum normalized velocity deficit as the element size is decreased. We observe that mesh sensitivity is sufficiently reduced at an element size of $1 D$ since further reduction to $0.5 D$ only results in a $0.69\,\si{\percent}$ change in the predicted maximum normalized velocity deficit. An element size of $1 D$ is also appropriate from the standpoint of time-efficiency as it requires $3.2\,\si{sec}$ of computation time to run a $5\,\si{sec}$ simulate.

\begin{table}
	\centering
	\caption{Computation times and maximum normalized velocity deficits corresponding to different simulated mesh element sizes from Fig.~\ref{Figure - Mesh convergence results}. The final column lists the convergence of the maximum normalized velocity deficit. In other words, it contains the relative difference in the maximum normalized velocity deficit that would be obtained if each element size was halved. For instance, if the element size were to be reduced from $8 D$ to $4 D$, the predicted velocity deficit would change by $19.46\,\si{\percent}$. The computation times correspond to $5\,\si{sec}$ long simulations.}
	\label{Table - Mesh convergence results}
	\scriptsize
	\begin{tabular}{|c|c|c|c|}
		\hline
		Elm. size (D) & Comp. time (sec) & Max. velocity deficit (-) & Rel. diff. (\%) \\ \hline \hline
		8 & 0.397 & 0.261 & 19.46 \\ \hline
		4 & 0.703 & 0.219 & 7.19 \\ \hline
		2 & 1.496 & 0.204 & 2.00 \\ \hline
		1 & 3.237 & 0.200 & 0.69 \\ \hline
		0.5 & 8.948 & 0.199 & 0.17 \\ \hline
		0.25 & 24.035 & 0.198 & - \\ \hline
	\end{tabular}
\end{table}

\subsection{Validation at steady-state} \label{Subsection - Validation}

\begin{figure}
	\centering
	\includegraphics[width=4.5in]{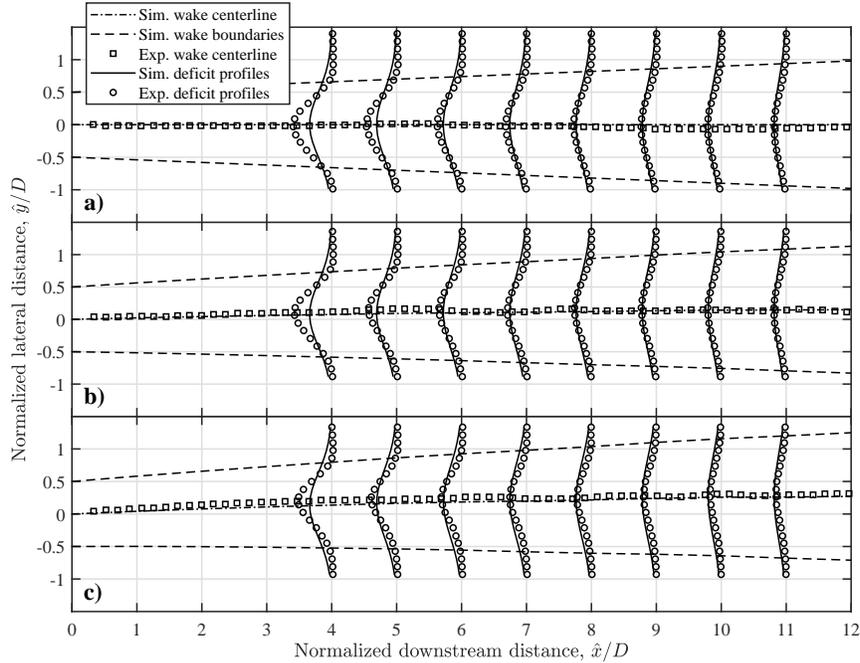}
	\caption{Comparison between FOWFSim-Dyn predictions and experimental results reported by Bastankhah and Port\'e Agel~\cite{Bastankhah2016}. Each figure shows steady-state wake centerlines and normalized velocity profiles corresponding to yaw angles of \textbf{a)} $\gamma = 0\,\si{\deg}$, \textbf{b)} $\gamma = 10\,\si{\deg}$, and \textbf{c)} $\gamma = 20\,\si{\deg}$. Normalized velocity profiles range from zero to one using the same scaling as the $\hat{x}/D$ axis, but have been shifted to the downstream location where they are measured. Simulation parameters: $D = 15\,\si{cm}$, $U_\infty = 8\,\si{m/s}$, $a = 1/3$, $k_x = 0.08$, $\sigma = 0.025 \hat{x} + 0.396\,\si{m}$.}
	\label{Figure - Validation results}
\end{figure}

FOWFSim-Dyn predictions of steady-state\footnote{Validating predictions of dynamic wake behaviour is not possible at this time due to the absence of high-fidelity simulation tools capable of modeling floating offshore wind farms; we thus defer this process to future work.} wake centerlines and normalized velocity profiles are compared against experimental results reported by Bastankhah and Port\'e Agel~\cite{Bastankhah2016} in Fig.~\ref{Figure - Validation results}. Wake centerline evolutions are well-predicted for all simulated yaw angles and downstream locations. For yaw angles of $\gamma = 0$, $10$, and $20\,\si{deg}$, maximum discrepancies between predicted wake centerlines and experimental measurements are $6.87$, $7.60$, and $8.19\,\si{\percent}$ of the rotor diameter, respectively.

Simulated normalized velocity profiles deviate significantly from experimental measurements at downstream locations closer than $7 D$. For instance, at a yaw angle of $\gamma = 0\,\si{deg}$, the root-mean-square error (RMSE) between experimental and predicted velocity profiles ranges from $12.4\,\si{\percent}$ of the free stream wind speed at $\hat{x} = 4 D$ to $4.7\,\si{\percent}$ at $\hat{x} = 7 D$. Such inaccuracies at close downstream distances are expected since FOWFSim-Dyn does not consider the inviscid nature of flow within the near-wake region. Beyond $\hat{x} = 7 D$, velocity profiles are well-predicted with RMSE values that remain below $3.87\,\si{\percent}$ of the free stream wind speed.

\subsection{Dynamic simulation} \label{Subsection - Dynamic simulation}

\begin{figure}
	\centering
	\includegraphics[width=4in]{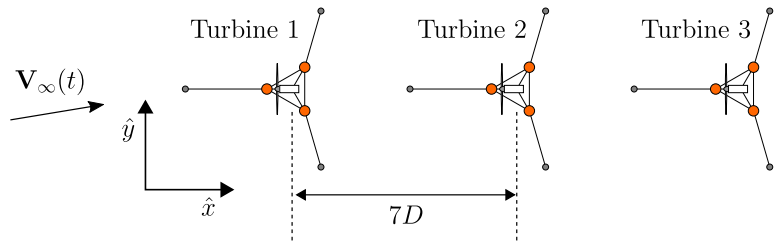}
	\caption{Schematic of the $1 \times 3$ wind farm with inter-turbine spacings of $7 D$ used for dynamic simulations. All wind turbines are based on the NREL $5\,\si{MW}$ baseline design presented by Jonkman \textit{et al.}~\cite{Jonkman2009}, and all floating platforms and mooring subsystems are modeled after the design described by Robertson \textit{et al.}~\cite{Robertson2014}.}
	\label{Figure - Simulated wind farm}
\end{figure}

Our final tasks are to demonstrate the capability of FOWFSim-Dyn to capture the intended dynamic phenomena and to ensure that predicted turbine and wake behaviors respect physical intuition. The wind farm configuration that is used for dynamic simulations is shown in Fig.~\ref{Figure - Simulated wind farm}. This plant contains a single row of three floating offshore wind turbines that are aligned with the predominant free stream wind direction. The neutral positions of the floating turbines are spaced $7 D$ apart. All wind turbines are based on the National Renewable Energy Laboratory's (NREL's) $5\,\si{MW}$ baseline design presented by Jonkman \textit{et al.}~\cite{Jonkman2009}, and all floating platforms and mooring subsystems are modeled after the design described by Robertson \textit{et al.}~\cite{Robertson2014}. Details corresponding to these designs are listed in Appendix~\ref{Appendix - Wind farm properties}. In all simulations, we increase the lengths of mooring lines from their baseline values (\textit{i.e.} $L = 835\,\si{m}$) to $L = 900\,\si{m}$ to render floating platform motion more notable. In all the following cases, less than $10\,\si{sec}$ of computation time was required to complete simulations on a laptop computer with a $2.80\,\si{GHz}$ Intel Core i7-7700HQ processor.

\subsubsection{Simulation scenario 1}

\begin{figure}
	\centering
	\includegraphics[width=5in]{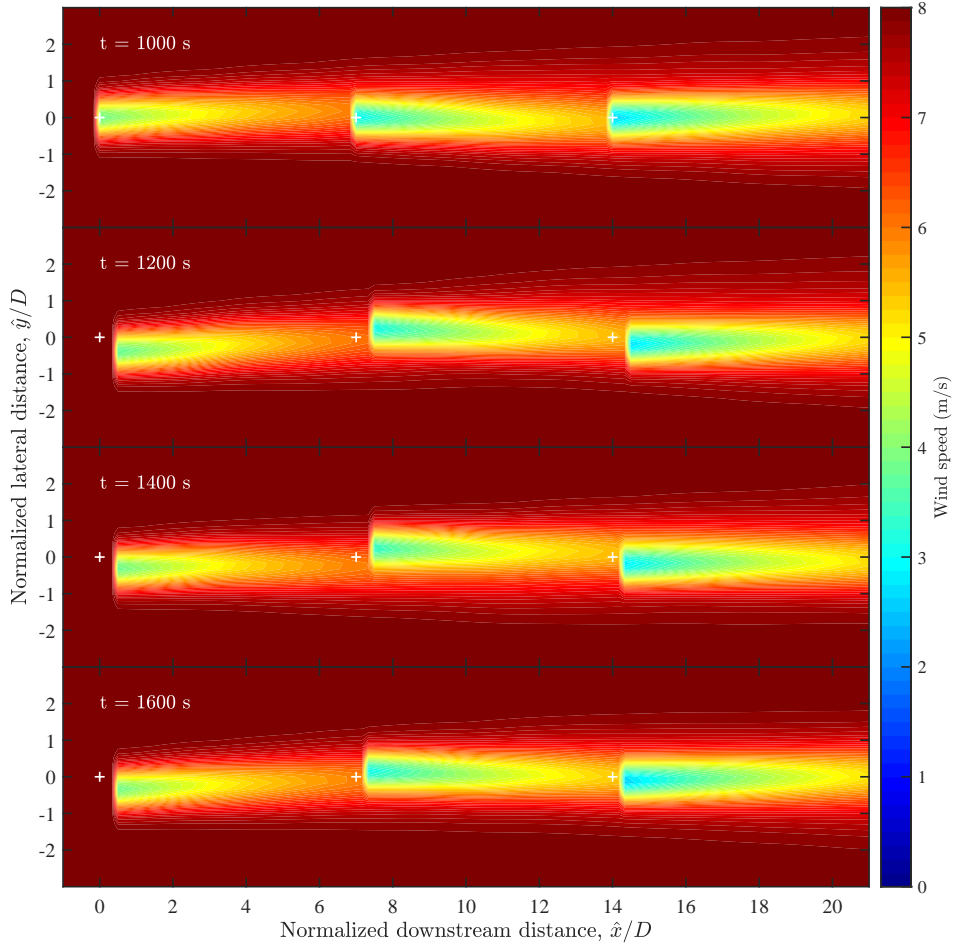}
	\caption{Velocity contours at various time-steps of simulation scenario~1 (\textit{i.e.} fixed wind condition and turbine operating parameters, while platform motion is permitted). The white $+$ symbols represent the neutral positions of the floating platforms. All floating platforms are held fixed at their respective neutral positions for the first $1000\,\si{sec}$ of simulation. Simulation parameters: $U_\infty(t) = 8\,\si{m/s}$, $V_\infty(t) = 0\,\si{m/s}$, $a_1(t) = a_2(t) = a_3(t) = 1/3$, $\gamma_1(t) = \gamma_3(t) = -20\,\si{deg}$ and $\gamma_2(t) = +20\,\si{deg}$, $k_x = 0.08$, $\sigma = 0.025 \hat{x} + 0.396\,\si{m}$.}
	\label{Figure - Dynamic simulation 1}
\end{figure}

The first of three simulated scenarios maintains constant wind speed and direction with $U_\infty(t) = 8\,\si{m/s}$ and $V_\infty(t) = 0\,\si{m/s}$, while rotor yaw angles are fixed at $\gamma_1(t) = \gamma_3(t) = -20\,\si{deg}$ and $\gamma_2(t) = +20\,\si{deg}$. All axial induction factors are maintained at $a_1(t) = a_2(t) = a_3(t) = 1/3$. All floating platforms are locked at their neutral positions for the first $1000\,\si{sec}$ of simulation, after which they are permitted to relocate. The aim of this scenario is to assess floating platform motion. Snapshots of velocity contours for simulation scenario~1 are shown in Fig.~\ref{Figure - Dynamic simulation 1}. As expected, the alternating assignment of yaw angles causes adjacent floating platforms to shift in opposite directions over time. Further, the leading turbine displays the greatest amount of relocation from its neutral position (\textit{i.e.} the left-most white $+$ symbol) since its incident wind speed is the largest (\textit{i.e.} its incident wind speed is the free stream wind speed uninhibited by upstream rotors). The trailing turbine undergoes the smallest amount of relocation over time since its incident wind speed is diminished by the velocity deficits of wakes~1 and~2.

\subsubsection{Simulation scenario 2}

\begin{figure}
	\centering
	\includegraphics[width=5in]{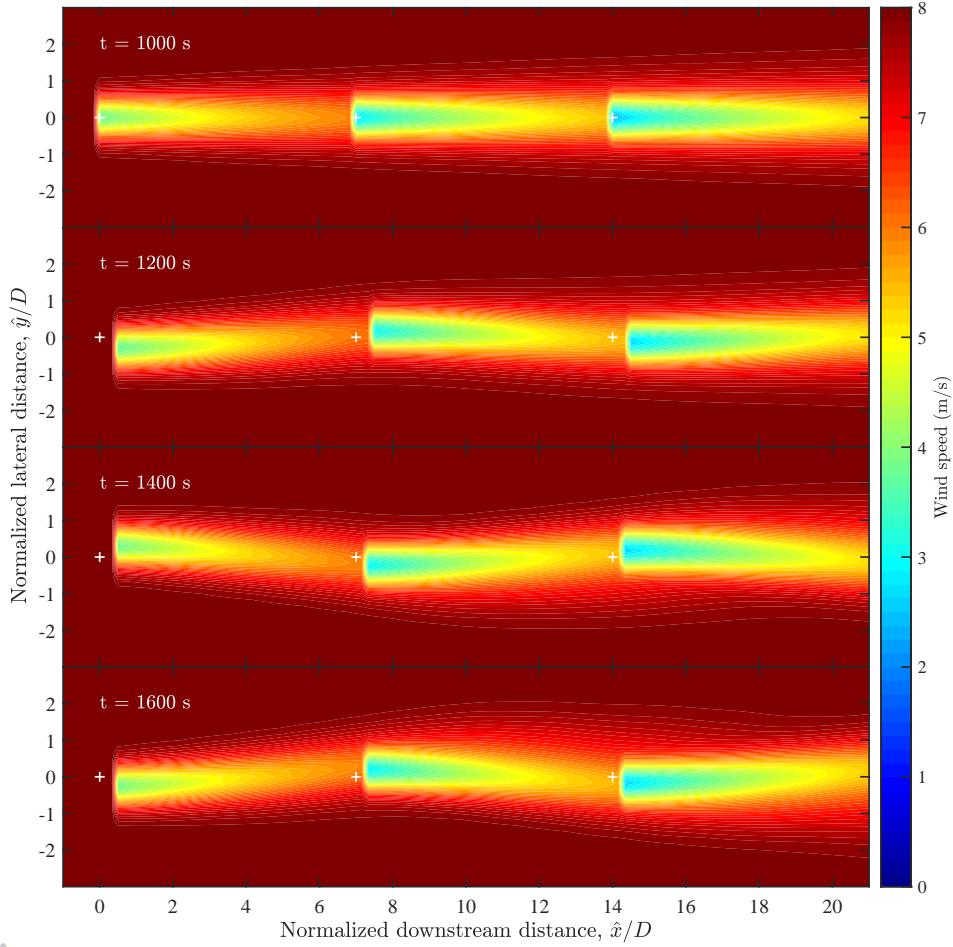}
	\caption{Velocity contours at various time-steps of simulation scenario~2 (\textit{i.e.} fixed wind condition and sinusoidally varying yaw angles, while platform motion is permitted). The white $+$ symbols represent the neutral positions of the floating platforms. All floating platforms are held fixed at their respective neutral positions for the first $1000\,\si{sec}$ of simulation. Simulation parameters: $U_\infty(t) = 8\,\si{m/s}$, $V_\infty(t) = 0\,\si{m/s}$, $a_1(t) = a_2(t) = a_3(t) = 1/3$, $\gamma_1(t)$ and $\gamma_3(t)$ defined in Eq.~(\ref{Equation - Yaw angles 1 and 3}) and $\gamma_2(t)$ defined in Eq.~(\ref{Equation - Yaw angle 2}), $k_x = 0.08$, $\sigma = 0.025 \hat{x} + 0.396\,\si{m}$.}
	\label{Figure - Dynamic simulation 2}
\end{figure}

The second simulation sinusoidally varies the yaw angles of the three turbines between $\pm 20\,\si{deg}$ with a period of $400\,\si{sec}$. Specifically, the following yaw angle expressions are used for $t \geq 1000\,\si{sec}$:
\begin{eqnarray}
	\gamma_1(t) = \gamma_3(t) & = & (-20\,\si{deg}) \sin \left[ \frac{2 \pi}{400} \left( t - 1000\,\si{sec} \right) \right], \label{Equation - Yaw angles 1 and 3} \\
	\gamma_2(t) & = & (+20\,\si{deg}) \sin \left[ \frac{2 \pi}{400} \left( t - 1000\,\si{sec} \right) \right].  \label{Equation - Yaw angle 2}
\end{eqnarray}
Velocity contours for this case are plotted in Fig.~\ref{Figure - Dynamic simulation 2}. The sinusoidal yaw angle fluctuations cause oscillations of floating platforms in the $\hat{y}$ direction with the expected $400\,\si{sec}$ excitation period. In terms of wake behaviour, the transport effect is clearly observed. As floating turbines shift in the $\hat{y}$ direction, the corresponding effects on their respective wakes are transported downstream at approximately $8\,\si{m/s}$. For instance, at $t = 1400\,\si{sec}$, the leading turbine is located at a peak value past its neutral position in the $+\hat{y}$ direction. Given that $U_\infty = 8\,\si{m/s}$ , then $200\,\si{sec}$ later, the centerline of the leading turbine's wake must peak in the $+\hat{y}$ direction at $\hat{x} = 8\,\si{m/s} \times 200\,\si{sec} = 1600\,\si{m} = 12.7 D$. Observing the velocity contours $200\,\si{sec}$ later at $t = 1600\,\si{sec}$, such a peak is observed at just under $\hat{x} = 12 D$.

\subsubsection{Simulation scenario 3}

\begin{figure}
	\centering
	\includegraphics[width=5in]{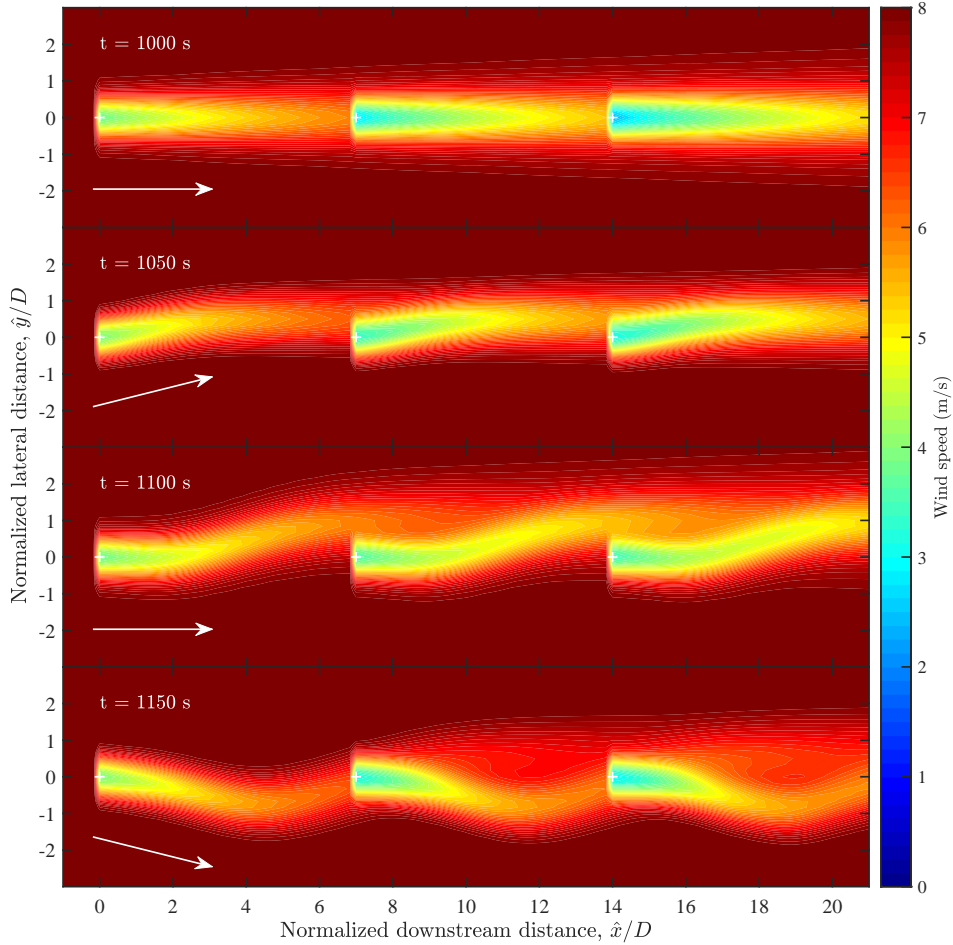}
	\caption{Velocity contours at various time-steps of simulation scenario~3 (\textit{i.e.} fixed turbine operating conditions and fluctuating wind speed in the $\hat{y}$ direction, while platform motion is prohibited). The white $+$ symbols represent the neutral positions of the floating platforms. The white arrows denote the free stream wind direction. All floating platforms are held fixed throughout the simulation. Simulation parameters: $U_\infty(t) = 8\,\si{m/s}$, $V_\infty(t)$ defined in Eq.~(\ref{Equation - Wind speed fluctuation}), $a_1(t) = a_2(t) = a_3(t) = 1/3$, $\gamma_1(t) = \gamma_2(t) = \gamma_3(t) = 0\,\si{deg}$, $k_x = 0.08$, $\sigma = 0.025 \hat{x} + 0.396\,\si{m}$.}
	\label{Figure - Dynamic simulation 3}
\end{figure}

The third scenario assesses the impacts of time-varying wind direction, which is modeled by maintaining $U_\infty(t) = 8\,\si{m/s}$ and fluctuating $V_\infty(t)$ sinusoidally between $\pm 2\,\si{m/s}$ with a period of $200\,\si{sec}$. Specifically, $V_\infty(t)$ is expressed as follows for $t \geq 1000\,\si{sec}$:
\begin{equation}
	V_\infty(t) =  (2\,\si{m/s}) \sin \left[ \frac{2 \pi}{200} \left( t - 1000\,\si{sec} \right) \right]. \label{Equation - Wind speed fluctuation}
\end{equation}
All yaw angles in this scenario are maintained at $\gamma_1(t) = \gamma_2(t) = \gamma_3(t) = 0\,\si{deg}$. Velocity contours for simulation case~3 are shown in Fig.~\ref{Figure - Dynamic simulation 3}. The notable expectation here is that, as the wind direction changes, wake centerlines are transported in tandem with the free stream wind in both $\hat{x}$ and $\hat{y}$ directions. For instance, at $t = 1000\,\si{sec}$, the centerline of wake~1 is aligned with the $\hat{x}$ axis since $\gamma_1(t) = 0\,\si{deg}$ and $V_\infty(t)$ had been equal to zero at all previous times. By $t = 1050\,\si{sec}$, the effects of turbine~1 on the wind field should only be transported downstream by a distance of $8\,\si{m/s} \times 50\,\si{sec} = 400\,\si{m} = 3.2 D$. Therefore, for $\hat{x} < 3.2 D$, we expect variations in the curvature of the centerline of wake~1 due to the presence of turbine~1, while for $\hat{x} > 3.2 D$, this curvature should remain unchanged. Instead, for $\hat{x} > 3.2 D$, the centerline of wake~1 should be shifted in the $+\hat{y}$ direction as a result of $V_\infty(t)$ having held positive values for the past $50\,\si{sec}$. Observing velocity contours at $t = 1050\,\si{sec}$, it is evident that the centerline curvature of wake~1 remains flat at all downstream distances past approximately $\hat{x} = 3 D$, while having been shifted in the $+\hat{y}$ direction.

\section{Conclusions and recommendations for future research} \label{Section - Conclusions and recommendations}

This paper extended FOWFSim~\cite{Kheirabadi2020}, which is a steady-state modeling tool that may be used for simulating and optimizing floating offshore wind farms, by adding capabilities that captured time-varying free stream wind velocities and floating platform motion. In addition to presenting a mathematical formulation, we performed a mesh convergence study and validated steady-state predictions on wake behaviour against experimental data obtained from existing literature. It was demonstrated that the limited number of tunable parameters produced wake centerline deflection and velocity deficit results that matched experimental observations with reasonable similarity for engineering analysis. We then conducted simulations under various wind and turbine operating conditions to assess the dynamic behavior of FOWFSim-Dyn. It was observed that FOWFSim-Dyn captures dynamic floating wind farm phenomena such as wake transport, time-varying wind speed and direction effects, and floating platform motion in line with physical reasoning and intuition.

For the purposes of further developing and enhancing the current framework, several recommendations on potential research directions are made. First, to this date, no LES-based wind farm simulators are capable of capturing floating platform motion. Developing wind farm CFD tools that consider such dynamics would therefore permit more comprehensive validation of FOWFSim-Dyn predictions pertaining to both platform motion and wake behaviour. Complementing this point, scaled wind tunnel experiments of floating wind turbines would also enable validation of dynamic FOWFSim-Dyn predictions.

Second, we did not model turbulence in the current framework. This feature may be incorporated by adding measurement noise to model outputs, or by including temporally and spatially distributed turbulence acceleration terms in the equations of motion. Finally, additional force gradients may be included in the equations of motion to capture complex wake phenomena such as secondary steering~\cite{Fleming2018} and wake deflection due to rotor rotation~\cite{Gebraad2016}.

\section*{Acknowledgment}

The authors are grateful for the financial support provided by the Natural Sciences and Engineering Research Council of Canada (NSERC).

\appendix

\section{Wind farm properties} \label{Appendix - Wind farm properties}

\begin{table}[H]
	\centering
	\caption{List of floating wind farm properties used during simulations that are discussed in Section~\ref{Section - Model tuning and validation}. All wind turbines are based on the NREL $5\,\si{MW}$ baseline design presented by Jonkman \textit{et al.}~\cite{Jonkman2009}, and all floating platforms and mooring subsystems are modeled after the design described by Robertson \textit{et al.}~\cite{Robertson2014}.}
	\label{Table - Wind farm properties}
	\scriptsize
	\begin{tabular}{|l|c|l|}
		\hline
		\multicolumn{3}{|l|}{External properties} \\ \hline
		$\rho_\mathrm{a}\,\left(\si{kg/m^3} \right)$ & 1.225 & Air density \\ \hline
		$\rho_\mathrm{w}\,\left(\si{kg/m^3} \right)$ & 1028 & Water density \\ \hline \hline
		\multicolumn{3}{|l|}{Floating turbine properties} \\ \hline
		$m_i\,\left(\si{kg}\right)$ & $1.4 \times 10^7$ & Mass \\ \hline
		$D_i\,\left(\si{m}\right)$ & $126$ & Rotor diameter \\ \hline
		$A_i\,\left(\si{m^2}\right)$ & $\frac{\pi}{4} D_i^2$ & Rotor area \\ \hline
		$\eta_\mathrm{p}$ & $0.786$ & Electrical power conversion efficiency~\cite{Gebraad2016} \\ \hline
		$p_\mathrm{p}$ & $1.88$ & Power coefficient tuning parameter~\cite{Gebraad2016} \\ \hline \hline
		\multicolumn{3}{|l|}{Floating platform hydrodynamic properties} \\ \hline
		$C_{\mathrm{d},i,1 \rightarrow 3}$ & $0.61$ & Drag coefficients of three top cylinder portions \\ \hline
		$C_{\mathrm{d},i,4 \rightarrow 6}$ & $0.68$ & Drag coefficients of three bottom cylinder portions \\ \hline
		$C_{\mathrm{d},i,7}$ & $0.56$ & Drag coefficient of middle cylinder \\ \hline
		$D_{\mathrm{d},i,1 \rightarrow 3}\,\left(\si{m}\right)$ & $12$ & Diameters of three top cylinder portions \\ \hline
		$D_{\mathrm{d},i,4 \rightarrow 6}\,\left(\si{m}\right)$ & $24$ & Diameters of three bottom cylinder portions \\ \hline
		$D_{\mathrm{d},i,7}\,\left(\si{m}\right)$ & $6.5$ & Diameter of middle cylinder \\ \hline
		$L_{\mathrm{d},i,1 \rightarrow 3}\,\left(\si{m}\right)$ & $14$ & Submerged lengths of three top cylinder portions \\ \hline
		$L_{\mathrm{d},i,4 \rightarrow 6}\,\left(\si{m}\right)$ & $6$ & Submerged lengths of three bottom cylinder portions \\ \hline
		$L_{\mathrm{d},i,7}\,\left(\si{m}\right)$ & $20$ & Submerged length of middle cylinder \\ \hline
		$A_{\mathrm{d},i,j}\,\left(\si{m^2}\right)$ & $L_{\mathrm{d},i,j} D_{\mathrm{d},i,j}$ & Drag reference area of any cylinder \\ \hline
		$C_{\mathrm{a},i,j}$ & $0.63$ & Added mass coefficients of any cylinder \\ \hline
		$A_{\mathrm{a},i,j}\,\left(\si{m^2}\right)$ & $\frac{\pi}{4}L_{\mathrm{d},i,j} D_{\mathrm{d},i,j}^2$ & Added mass reference area of any cylinder \\ \hline \hline
		\multicolumn{3}{|l|}{Mooring system properties} \\ \hline
		$\mathbf{r}_{\mathrm{F}/\mathrm{G},i,1}^\mathrm{T}\,\left(\si{m}\right)$ & $\begin{bmatrix} 20.4 & 35.4 \end{bmatrix}$ & Position vector from turbine center to first fairlead \\ \hline
		$\mathbf{r}_{\mathrm{F}/\mathrm{G},i,2}^\mathrm{T}\,\left(\si{m}\right)$ & $\begin{bmatrix} -40.9 & 0 \end{bmatrix}$ & Position vector from turbine center to second fairlead \\ \hline
		$\mathbf{r}_{\mathrm{F}/\mathrm{G},i,3}^\mathrm{T}\,\left(\si{m}\right)$ & $\begin{bmatrix} 20.4 & -35.4 \end{bmatrix}$ & Position vector from turbine center to third fairlead \\ \hline
		$\mathbf{r}_{\mathrm{A},i,1}^\mathrm{T}\,\left(\si{m}\right)$ & $\mathbf{r}_{\mathrm{neutral},i}^\mathrm{T} + \begin{bmatrix} 418.80 & 725.4 \end{bmatrix}$ & Location of first anchor of any turbine\tablefootnote{Anchors are located at angles of $60$, $180$, and $300\,\si{deg}$ along a circle of radius $837.6\,\si{m}$ surrounding the neutral positions $\mathbf{r}_{\mathrm{neutral},i}$ of their respective turbines.} \\ \hline
		$\mathbf{r}_{\mathrm{A},i,1}^\mathrm{T}\,\left(\si{m}\right)$ & $\mathbf{r}_{\mathrm{neutral},i}^\mathrm{T} + \begin{bmatrix} -837.6 & 0 \end{bmatrix}$ & Location of second anchor of any turbine \\ \hline
		$\mathbf{r}_{\mathrm{A},i,1}^\mathrm{T}\,\left(\si{m}\right)$ & $\mathbf{r}_{\mathrm{neutral},i}^\mathrm{T} + \begin{bmatrix} 418.80 & -725.4 \end{bmatrix}$ & Location of third anchor of any turbine \\ \hline
		$z_\mathrm{F}\,\left(\si{m}\right)$ & $186$ & Fairlead distance above seabed \\ \hline
		$L\,\left(\si{m}\right)$ & $835$ & Cable length\tablefootnote{Simulations corresponding to Figs.~\ref{Figure - Dynamic simulation 1} to \ref{Figure - Dynamic simulation 3} use longer cable lengths of $L = 900\,\si{m}$.} \\ \hline
		$w\,\left(\si{N/m}\right)$ & $1065.7$ & Cable weight per unit length in water \\ \hline
		$A_\mathrm{m} E\,\left(\si{N}\right)$ & $753.6 \times 10^6$ & Cable tension per unit strain \\ \hline
		$\mu_\mathrm{s}$ & $1$ & Coefficient of static friction between cable and seabed \\ \hline
	\end{tabular}
\end{table}

\section{Formulae for computing mooring line tension} \label{Appendix - Catenary solution}

This appendix section briefly details the formulae used to calculate the horizontal component of tension within any mooring line cable. Derivations of the following formulae may be found in our previous work~\cite{Kheirabadi2020}. For readability, we drop functional time-dependency indicators and subscripts (\textit{i.e.} $H_{\mathrm{F},i,k}(t)$ simply becomes $H_\mathrm{F}$) since the discussed solution is static and all formulae remain the same for any individual mooring line cable.

We begin by defining three zones of mooring line operation. The first zone is in effect when the fairlead is close enough to its respective anchor that the cable is vertical at the fairlead location; the resulting horizontal component of tension is zero in this case. The second zone of operation occurs when the cable is partially contacting the seabed, while the third zone is relevant when the cable is fully lifted off of the seabed. Based on these definitions, we define $H_\mathrm{F}$ as follows:
\begin{equation}
\label{Equation - Cable tension}
H_\mathrm{F} = \left\{
\begin{array}{lll}
0 & \mathrm{if} & x_\mathrm{F} \leq x_{\mathrm{F},1 \rightarrow 2}, \\
f_1 & \mathrm{if} & x_{\mathrm{F},1 \rightarrow 2} < x_\mathrm{F} \leq x_{\mathrm{F},2 \rightarrow 3}, \\
f_2 & \mathrm{if} & x_\mathrm{F} \geq x_{\mathrm{F},2 \rightarrow 3},
\end{array}
\right.
\end{equation}
where $x_\mathrm{F}$ is the horizontal distance from the fairlead to its respective anchor as follows:
\begin{equation}
x_\mathrm{F} = \left\Vert \mathbf{r}_{\mathrm{F}/\mathrm{A},i,k} \right\Vert,
\end{equation}
and the fairlead locations of transition between the different zones are computed as follows:
\begin{eqnarray}
x_{\mathrm{F},1 \rightarrow 2} & = & L - z_\mathrm{F}, \\
x_{\mathrm{F},2 \rightarrow 3} & = & \frac{H_{2 \rightarrow 3}}{w} \left[ \frac{w L}{A_\mathrm{m} E} + \sinh^{-1} \frac{w L}{H_{2 \rightarrow 3}} \right].
\end{eqnarray}
The parameters $L$, $w$, $A_\mathrm{m}$, and $E$ represent the length, specific weight in water, cross-sectional area, and elastic modulus of the cable, $z_\mathrm{F}$ is the vertical distance between the fairlead and its respective anchor, and $H_{2 \rightarrow 3}$ is the horizontal tension within the cable at the transition between operating zones~2 and~3, which has been derived to give the following expression:
\begin{equation}
H_{2 \rightarrow 3} = \frac{w L}{2} \left[ 1 - \left( \frac{z_\mathrm{F}}{L} - \frac{w L}{2 A_\mathrm{m} E} \right)^2 \right] \left( \frac{z_\mathrm{F}}{L} - \frac{w L}{2 A_\mathrm{m} E} \right)^{-1}.
\end{equation}

The function $f_1$ from Eq.~(\ref{Equation - Cable tension}) solves the following system of nonlinear equations for the horizontal and vertical components of cable tension $H_\mathrm{F}$ and $V_\mathrm{F}$:
\begin{eqnarray}
x_\mathrm{F} - L_\mathrm{s} & = & \frac{H_\mathrm{F}}{w} \left( \frac{V_\mathrm{F}}{A_\mathrm{m} E} + \sinh^{-1} \frac{V_\mathrm{F}}{H_\mathrm{F}} \right), \\
z_\mathrm{F} & = & \frac{1}{w} \left\{ \frac{V_\mathrm{F}^2}{2 A_\mathrm{m} E} - H_\mathrm{F} \left[ 1 - \sqrt{1 + \left( \frac{V_\mathrm{F}}{H_\mathrm{F}} \right)^2} \right] \right\}.
\end{eqnarray}
These equations correspond to a catenary profile that is partially contacting the seabed. The term $L_\mathrm{s}$ is the length of the cable portion that is contacting the seabed, which we derive to yield the following expression:
\begin{equation}
L_\mathrm{s} = L - \frac{V_\mathrm{F}}{w} + \frac{\left( 1 + \frac{H_\mathrm{F}}{A_\mathrm{m} E} \right)^3 - \left[ \left( 1 + \frac{H_\mathrm{F}}{A_\mathrm{m} E} \right)^2 - \frac{2 \mu_\mathrm{s} w}{A_\mathrm{m} E} x_\mathrm{s} \right]^{\frac{3}{2}}}{\frac{3 \mu_\mathrm{s} w}{A_\mathrm{m} E}} - x_\mathrm{s}.
\end{equation}
The term $x_\mathrm{s}$ represents the location along the seabed-contacting portion at which the total static friction force equates the cable tension. Our derivation for $x_\mathrm{s}$ is expressed as follows:
\begin{equation}
x_\mathrm{s} = \min \left[ L - \frac{V_\mathrm{F}}{w}, \frac{H_\mathrm{F}}{\mu_\mathrm{s} w} \left( 1 + \frac{H_\mathrm{F}}{2 A_\mathrm{m} E} \right) \right].
\end{equation}

Similarly, the function $f_2$ from Eq.~(\ref{Equation - Cable tension}) solves the following system of nonlinear equations for the horizontal and vertical components of cable tension $H_\mathrm{F}$ and $V_\mathrm{F}$:
\begin{eqnarray}
x_\mathrm{F} & = & \frac{H_\mathrm{F}}{w} \left( \frac{w L}{A_\mathrm{m} E} + \sinh^{-1} \frac{V_\mathrm{F}}{H_\mathrm{F}} - \sinh^{-1} \frac{V_\mathrm{F} - w L}{H_\mathrm{F}} \right), \\
z_\mathrm{F} & = & \frac{L}{A_\mathrm{m} E} \left( V_\mathrm{F} - \frac{w L}{2} \right) + \frac{H_\mathrm{F}}{w} \left[ \sqrt{1 + \left( \frac{V_\mathrm{F}}{H_\mathrm{F}} \right)^2} - \sqrt{1 + \left( \frac{V_\mathrm{F} - w L}{H_\mathrm{F}} \right)^2} \right].
\end{eqnarray}
These equations correspond to a catenary profile that is fully-lifted off of the seabed.

\bibliographystyle{unsrt}
\bibliography{Library.bib}

\end{document}